# Spectral compression of single-photon-level laser pulse

**Yuanhua Li,[1,2] Tong Xiang,[1,2] Yiyou Nie,[3] Minghuang Sang,[3] and Xianfeng Chen[1,2,*]**

[1]State Key Laboratory of Advanced Optical Communication Systems and Networks, Department of Physics and Astronomy, Shanghai Jiao Tong University, Shanghai 200240, China

[2]Key Laboratory for Laser plasma (Ministry of Education), Collaborative Innovation Center of IFSA (CICIFSA), Shanghai Jiao Tong University, Shanghai 200240, China

[3]Department of Physics, Jiangxi Normal University, Nanchang 330022, China

*xfchen@sjtu.edu.cn

**ABSTRACT**
We experimentally demonstrate that the bandwidth of single photons laser pulse is compressed by a factor of 58 in a periodically poled lithium niobate (PPLN) waveguide chip. A chirped single photons laser pulse and an oppositely chirped classical laser pulse are employed to produce a narrowband single photon pulse with new frequency through sum-frequency generation. In our experiment, the frequency and bandwidth of single photons at 1550 nm are simultaneously converted. Our results mark a critical step towards the realization of coherent photonic interface between quantum communication at 1550 nm and quantum memory in the near-visible window.

## 1.  Introduction

Low loss transmission is an intrinsic and unique property for single photons at 1550 nm in optical fiber.[1-3] In quantum networks over optical fiber, single photons at 1550 nm are used for virtually all quantum information tasks, such as quantum metrology,[4] quantum computation[5] and quantum cryptography.[6] Spontaneous parametric down-conversion (SPDC) sources are readily available for the production of entangled photon pairs at 1550 nm, and typically yield spectral bandwidths of 300 GHz.[7] Nevertheless, the narrowband photons in the near-visible wavelength possess the most efficient quantum memories and an ability of being easily detected by a silicon avalanche photodiode (APD). Therefore, it is highly expected that a coherent photonic interface is necessary which is capable of spectrum compressing and frequency conversing in the telecom band simultaneously.

Numerous schemes for spectral compression of broadband classical light have been demonstrated by using various compressor units and second-order nonlinear crystals.[8-23] Nonlinear frequency conversion could be used to generate the long narrowband pulse light,[24-29] such as sum frequency generation (SFG) in periodically poled lithium niobate (PPLN) waveguide.[30-32] Recently, a theoretical scheme has been proposed for spectral compression and frequency conversion of quantum light pulse.[33] The related experiment has been demonstrated that the spectrum of single photons was compressed by a factor of 40 in a β-barium-borate (BBO) crystal.[34] Similar results have been achieved through temporal gating[35] and room-temperature diamond quantum memory.[36] In the proposed methods of Ref. [34-36], spectrum compressing and frequency conversing technologies are not in 1550-nm telecommunications band, which is not appropriate for conventional quantum communication using all-optical-fiber networks. It is well known that single photons coherent state at 1550 nm is important in the development of future quantum communication tasks such as quantum key distribution.[3] To the best of our knowledge, spectral compression of single photons coherent state at 1550 nm has not been experimentally demonstrated yet.

In this Letter, we exploit a SFG process with an oppositely chirped classical laser pulse and a chirped single photons laser pulse, which the bandwidth of the chirped single photons laser pulse is compressed in a PPLN waveguide chip—from 800 GHz to 13.7 GHz—which is approaching the bandwidth regime of some quantum memories.[37,38] In the same time, the 1550-nm telecom-band photons are flexibly converted into the near infrared window.

## 2. Experimental results and discussion

In our experiment, spectral compressed single photons coherent pulse is generated by SFG between an antichirped classical laser pulse and a chirped single photons laser pulse. The oppositely chirped classical laser pulse of frequency with $\nu_{0,P}$ increases in time $t_1$, i.e., $\nu_{0,P}(t_1) = \nu_{0,P} + \pi t_1/A$, and the chirped single photons pulse frequency with $\nu_{0,Q}$ decreases linearly in time $t_2$, $\nu_{0,Q}(t_2) = \nu_{0,Q} - \pi t_2/A$. Here $A$ denotes the chirp rate, $\nu_{0,P}$ and $\nu_{0,Q}$ are the centre frequency of these two sources, and $\Delta t = t_1 - t_2$ is a relative time delay between the oppositely chirped classical laser pulse and chirped single photons laser pulse. When a chirped single photons laser pulse and a classical antichirped laser pulse arrive at the PPLN waveguide chip simultaneously, the frequency of $\nu_{0,Q}(t_2)$ would meet the frequency of $\nu_{0,P}(t_1)$ with the relative time delay $\Delta t$, and SFG happen between two pluses, and the created single photon pulse with a narrow frequency bandwidth can be described $\nu_{0,SFG}(\Delta t) = \nu_{0,P} + \nu_{0,Q} + \pi \Delta t/A$. $\nu_{0,Q}$ and $\nu_{0,P}$ are equal from two replicas of the laser pulse source as the chirped single photons laser pulse and the oppositely chirped classical laser pulse. The expected (TH) intensity bandwidth (FWHM) of the up-converted single photon is [34]

$$\Delta \nu_{SFG}^{TH} \approx \frac{\ln 4}{A} \sqrt{\frac{1}{\Delta \nu_P^2} + \frac{1}{\Delta \nu_Q^2}} \;, \tag{1}$$

where $\Delta \nu_P$ and $\Delta \nu_Q$ are the FWHM bandwidths used in the waveguide. The bandwidth $\Delta \nu$ is limited by group velocity dispersion, and decreases linearly with the length of the waveguide, i.e., $\Delta \nu = \Delta \hat{\nu}/L$, where $\Delta \hat{\nu} > 4200 GHz \bullet cm$ is the spectral acceptance of the waveguide. When the full FWHM bandwidths of the antichirped and chirped photons are used in the waveguide, the maximum SFG efficiency is guaranteed.

The experimental setup is shown in Fig. 1. Two variable optical attenuators (ATT1 and ATT2) are used to create single photons pulse and control the energy of an antichirped classical laser pulse for this experiment. The attenuation can also be realized by a polarization controller and a single mode polarization beam splitter. In our experiment, the laser pulse source has 500-fs duration, about 6.4 nm spectral FWHM bandwidth, 1551.54 nm centre wavelength at a 59.98 MHz repetition rate, and 45.2 mw average power.

Firstly, the laser pulse source is divided into two replicas using a 50:50 polarization maintaining beam splitter (BS), and then one of the two replicas of laser pulse is sent to a broadband fiber Bragg grating 1 (FBG1), and at the same time, the other laser pulse is coupled into the FBG2. The parameters of FBG1 and FBG2 are the exactly same (1547 nm centre wavelength, 39 nm FWHM bandwidth, and 5 nm/cm chirp rate). As ones know, FBG can be used for up-chirping and down-chirping, depending on the choice of the side from which the laser pulse is reflected. Thus, the two different chirp laser pulses after FBG1 and FBG2 are the same but with the opposite sign. It implies that the chirped and antichirped chirp pulses have equal and opposite chirp, $\pm A$. A chirped laser pulse is generated through FBG2 to introduce a linear chirp by group velocity dispersion, and the chirped single photons laser pulse is realized using attenuator 2 (ATT2) in our scheme. The other laser pulse is antichirped after a broadband FBG1.

The chirped single photons laser pulse and the oppositely chirped classical laser pulse are combined by a 1550 nm/1550 nm single mode beam splitter (SBS (50:50)) and coupled into the z-cut PPLN waveguide chip by the fiber pigtail. Two polarization controllers (PC) and two 200:1 single mode polarization beam splitter (SPBS) are used for adjusting the chirped single photons laser pulse and oppositely chirped classical laser pulse to the TM mode, that it supports Type-0 ($ee \rightarrow e$) phase matching in our experiment for the PPLN waveguide chip. A stable temperature controller is used to keep the PPLN waveguide chip's temperature to maintain the phase-matching of the SFG process. The spectrally narrowed single photon pulse of higher frequency is generated in a PPLN waveguide chip after interference filter (IF), with a nominal bandwidth of 20 nm (FWHM) centred around 780 nm (loss is about 1.2 dB), and coupled to an optical-fiber-coupled spectrometer. Finally, the SFG photons are detected by a Silicon APD (SAPD), whose detection efficiency is up to 60% at 775 nm and dark count rate is 26 cps. In our experiment, a superconducting single photon detector (SSPD) is used to calibrate and monitor the counts of the chirped single photons, whose detection efficiency is up to 10% at 1551 nm and dark count rate is 600 cps. Any residue of the chirped or antichirped light has to be filtered out from the SFG photons by a factor of $10^{-18}$.

The 5.2 cm long reverse-proton-exchange PPLN waveguide chip is quasi–phase matched to perform the SFG process 1551 nm+1551 nm→775.5 nm. It is poled with a quasi-phase-matching (QPM) period of 19.6μm, which incorporated single mode filters designed to match the single mode size of SMF-28 optical fiber and is optical fiber pigtailed with low coupling losses of about 0.7 dB. The PPLN waveguide chip has a total fiber-to-output-facet throughput of approximately −1.5 dB for the 1550-nm telecommunications band.

We first measure the spectrum of the chirped laser pulse by using an optical-fiber-coupled spectrometer and find a width $800 \pm 20$ GHz FWHM centred around 1551.54 nm. The chirped laser pulse is then sent into an optical fiber and superposed with the oppositely chirped classical laser pulse ($790 \pm 20$ GHz, centred around 1551.54 nm) in the PPLN waveguide chip for SFG. Here the FWHM bandwidths of both the chirped and antichirped photons are smaller than the spectral acceptance of the waveguide ($\Delta v = \Delta \hat{v}/L > 807$ GHz). Thus, the full FWHM bandwidths of the chirped and antichirped photons are used in the waveguide, as expected. The up-converted laser pulse generated by the SFG process, after IF, is coupled into a single-mode optical fiber and coupled into the spectrometer. As shown in Fig. 2 (a), we observe significant bandwidth compression of the chirped laser pulse. The measured bandwidth of the created laser pulse is $\Delta v_M = 33 \pm 1$ GHz (FWHM), centred at 775.77 nm, where the relative time delay $\Delta t = 0$. Taking the resolution of our spectrometer into account, $\Delta v_R = 30 \pm 1$ GHz (FWHM), the actual bandwidths of the up-converted photons after deconvolution is $\Delta v_{SFG}^{EXP} = 13.7 \pm 4.2$ GHz (FWHM). This result agrees closely with theory, $\Delta v_{SFG}^{TH} = 9.8 \pm 0.7$ GHz (FWHM) from equation (1), using the expected chirp parameter $A = (-2.52 \pm 0.01) * 10^8 \, fs^2$ given by the geometry of our FBG. Therefore, a spectral compression ratio of 58:1 is realized in the chirped laser pulse frequency bandwidth (Fig. 2 (b)).

The centre wavelength of the narrowband up-converted laser pulse can be tuned by adjusting the relative delay $\Delta t$ between the oppositely chirped laser pulse and the chirped laser pulse. The SFG spectrum of the created laser pulse could be given by a function of the delay, with the fitted centre wavelengths shown in Fig. 3. The experimental results show that the wavelength depends linearly on the delay, as expected. The linear fit gives a slope of -0.0247±0.001 nm/ps. In terms of the slope data, we obtain the oppositely chirp parameter of $A = (-2.55 \pm 0.01) * 10^8 \, fs^2$, in good agreement with the chirp parameter $A = (-2.52 \pm 0.01) * 10^8 \, fs^2$ of the FBG1. It is also found that the spectral compression ratio independent of the optical relative delay $\Delta t$, which agrees closely with the theoretical result from the above equation (1).

However, in our experiment, the up-converted single photons pulse light consists of second harmonic generation (SHG) of oppositely chirped laser pulse, SHG of chirped single photons laser pulse, and SFG of oppositely chirped laser pulse and chirped single photons laser pulse. In the experiment, we first couple the oppositely chirped laser pulse into the PPLN waveguide chip alone, and we measure the photons $P_1$ of SHG of oppositely chirped laser pulse. Similarly, the photons $P_2$ of SHG of chirped single photons laser pulse are measured when we only subject the chirped single photons laser pulse to the PPLN waveguide chip alone. If we simultaneously send the oppositely chirped laser pulse and chirped single photons laser pulse together to the PPLN waveguide chip, we obtain the photons $P_0$ of SFG and SHG. When the number of photons per pulse of chirped light is attenuated to single-photon level, the detected SHG counts drop to its dark counts (3.5 Hz). If the input energy of the antichirped laser pulse is 0.6 nJ, the $P_1$ is equal to the dark counts. In the case, it is verified that any photon detected by the SAPD is the result of the SFG process, and not SHG of the chirped and antichirped photons. Therefore, the photons per second of SFG are calculated according to the equation $P_{SFG} = P_0$. The SFG efficiency is then given by $\eta_{SFG} = \frac{P_{SFG}}{\beta N}$, where $\beta$ is the laser repetition rate, and $N$ is average of photons per second of chirped laser pulse.

As shown in Fig. 4, the energy of the antichirped laser pulse that is incident upon the PPLN waveguide chip is 0.6 nJ. By controlling the ATT2, the average of photons per second ($N_1 = 0.933$ and $N_2 = 0.302$) of chirped laser pulse can be obtained. SFG photons and SFG efficiencies of different situations are obtained by adjusting the relative delay $\Delta t$ (like Fig. 3 (b)). At the same time, it is found that the overall conversion efficiency of SFG varies with the relative delay $\Delta t$. When the relative time delay $\Delta t = 0$, the biggest efficiency of the SFG is $7.82 \times 10^{-6}$ with the average of photons of chirped laser pulse (0.933). Here the total losses have been considered, such as the coupling loss of 0.7 dB, total fiber-to-output-facet loss of 1.5 dB, reflection loss of 1.2 dB, and detection efficiency of 60% (see Fig.4).

The lower SFG signal for single photons required longer times than for the intense photons states (to reduce the effects of drift in experimental parameters all the data in Fig. 4 are taken within a day). The results show that the average of photons of the chirped pulse are more, the efficiency of SFG is larger (see Fig.4).

Furthermore, the SFG efficiency will decrease with reducing the antichirped photons. It is also found that the SFG efficiency increases with increasing input energy of the chirped and antichirped light. Next, we measure the SFG efficiency in two ways: one, by attenuating the oppositely chirped photons with the ATT1 and ATT2; the other by increasing the power of the chirped and antichirped laser with the ATT1 and ATT2. Figure 5 depicts the results of these two measurements.

As shown in Fig. 5 (a), when the oppositely chirped photons (ten photons per pulse) and chirped single photons laser pulse (0.933 photons per pulse) are simultaneously sent to the waveguide, the SFG efficiency of $4.58\times10^{-7}$ is realized, where the relative time delay $\Delta t = 0$. As shown in Fig. 5 (b), by adjusting the ATT1 and ATT2, we control the input energy of the chirped laser pulse and antichirped laser pulse is 203.1μJ and 202.8μJ, respectively. The energy of produced harmonics $E_i(i=0,1,2)$ is measured, where $E_0$ is the total energy of SFG and SHG of the chirped laser pulse and oppositely chirped laser pulse, $E_1$ is the energy of SHG of the chirped laser pulse, $E_2$ is the energy of SHG of the oppositely chirped laser pulse. When the relative time delay $\Delta t = 0$, the energy of produced harmonics is $E_0 = 21.62\ \mu J$, which is obtained from SHG of chirped laser ($E_1 = 0.28\ \mu J$), SHG of antichirped laser ($E_2 = 0.01\ \mu J$), and SFG ($E_{SFG} = E_0 - E_1 - E_2 = 21.33\ \mu J$). In the case, 20% overall conversion efficiency of SFG is obtained, where the total losses have been taken into account. We note that the rate of generated photons by SFG is 73 times of the rate of photons generated by the SHG of these two independent sources.

Our results may provide potential application in standard decoy-state quantum key distribution. By considering a fiber attenuation of 0.2 dB/km, the dark counts of 3.5 Hz, the coupling loss of 0.7 dB, total fiber-to-output-facet loss of 1.5 dB, reflection loss of 1.2 dB, detection efficiency of 60%, and sources with a 59.98 MHz repetition rate, the SFG efficiency of $7.82\times10^{-6}$ will achieve a rate of about 8 bits/hour on a distance of 20 km.

## 3. Conclusions

In conclusion, we have experimentally demonstrated that the spectrum of single photons laser pulse was compressed by a factor of 58 in a PPLN waveguide chip. A chirped single photons laser pulse and an oppositely chirped classical laser pulse by fiber Bragg gratings were used to produce a narrow bandwidth single photon pulse with new frequency through SFG process. The frequency and bandwidth of single photons at 1550 nm were simultaneously converted. Our results have demonstrated the potentially application for PPLN waveguide chip as an integrated platform for spectrum compressing and frequency conversing in the telecom band, such as coherent photonic interfaces between quantum communication at 1550 nm and quantum memory in the near-visible window.

**Acknowledgements**
This work was supported in part by the National Natural Science Foundation of China under Grant Nos. 61125503, 61235009, and 11564018, the Foundation for Development of Science and Technology of Shanghai under Grant No. 13JC1408300.

**Author contributions statement**
X. C. and Y. L. conceived the idea of this paper, Y. L. performed the simulations, drawn all of the figures and wrote the first draft, according to the experiment, design and data. T. X., Y. N., and M. S. provided experimental assistance. X. C. supervised the overall project. All authors discussed the results and reviewed the manuscript.

**Additional information**
Competing financial interests: The authors declare no competing financial interests.

**Figure Legends**

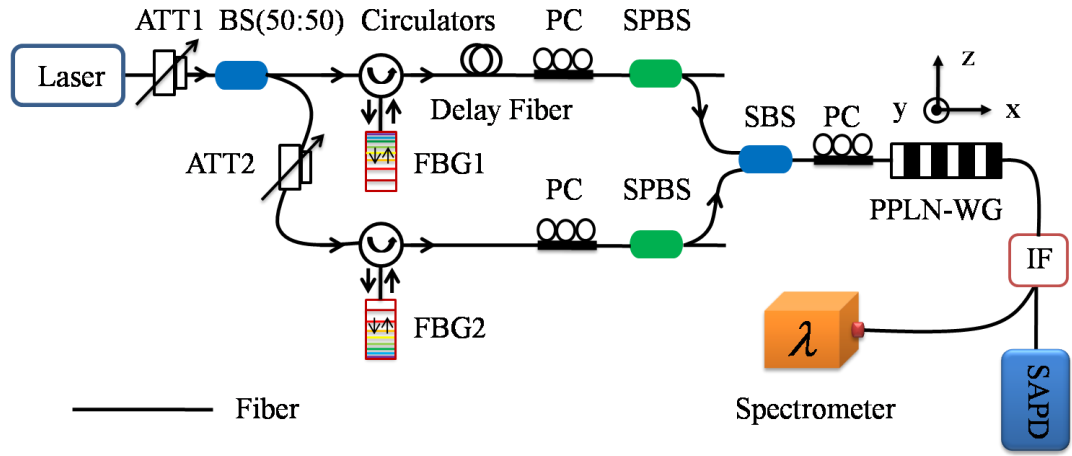

**Fig. 1** (color online). Experiment set-up. ATT, variable optical attenuator; BS, polarization maintaining beam splitter (50:50); FBG, fiber Bragg grating; Circulators, optical fiber circulators; Delay Fiber, optical adjustable delay fiber; PC, polarization controller; SPBS, single mode polarization beam splitter (single mode to polarization maintaining); SBS, single mode beam splitter (single mode to polarization maintaining); PPLN-WG, periodically poled lithium niobate waveguide chip; IF, interference filter; SAPD, Silicon APD.

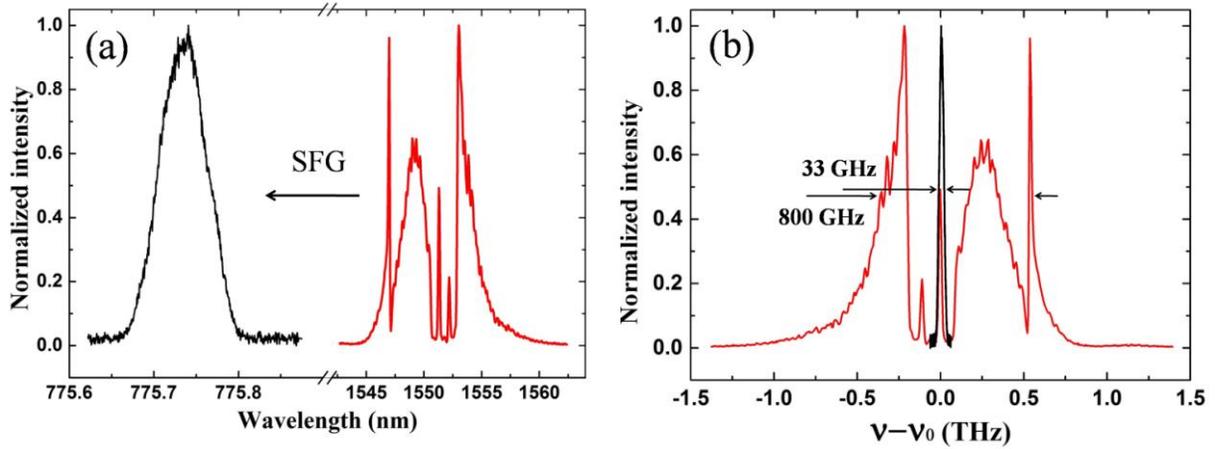

**Fig. 2** (color online). Chirped pulse spectrum and up-converted laser pulse spectrum (a) and relative frequency (b). The initial bandwidth of the chirped laser pulse is 800 GHz centred at 1551.54 nm (shown in red). Once the quadratic phase is applied and the two laser pulses are up-converted, the created laser pulse bandwidth reduces to $33\pm1$ GHz centred at 775.77 nm (shown in black). The spectra are given by normalized intensities and, for the up-converted case, correspond to the average of ten consecutive scans of 15 min acquisition time.

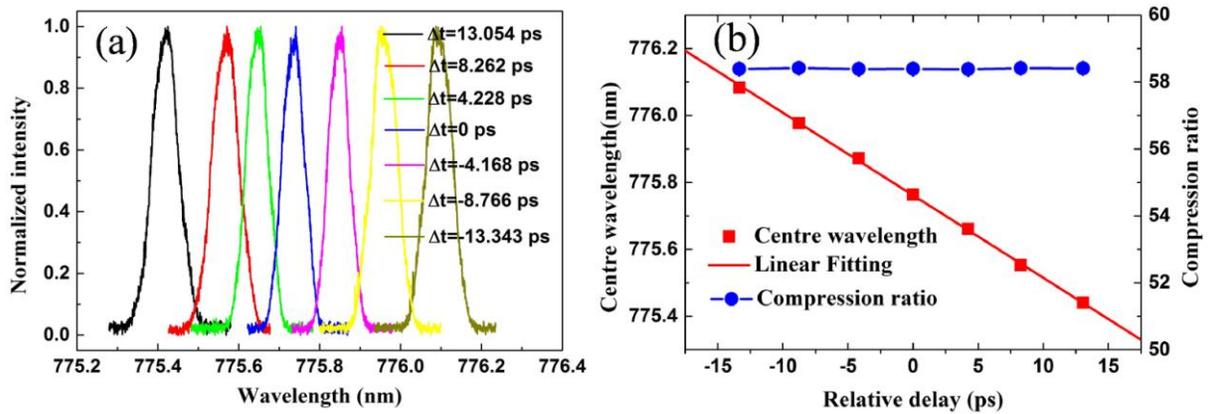

**Fig. 3** (color online). The SFG spectrum of the created laser pulse (a), central wavelength and compression ratio of the output pulses versus the optical relative delay (b). The central wavelength of the up-converted laser pulse is tuned by controlling the relative delay between the input pulses at the PPLN waveguide chip. Error bars are smaller than the data points.

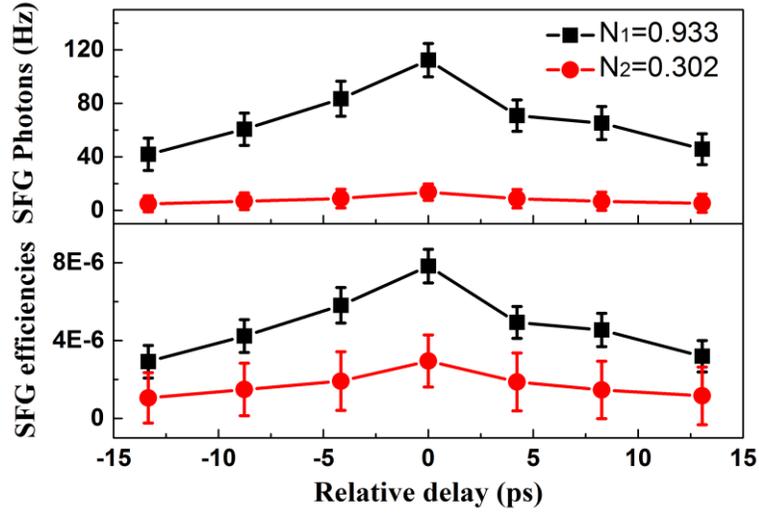

**Fig. 4** (color online). SFG photons (top) and SFG efficiencies (bottom). The dark counts (3.5 Hz) are subtracted. The SFG photons, SFG efficiencies and error bars are accounted and the abscissa is a variable optical relative delay between the oppositely chirped laser pulse and the chirped laser pulse at the PPLN waveguide chip.

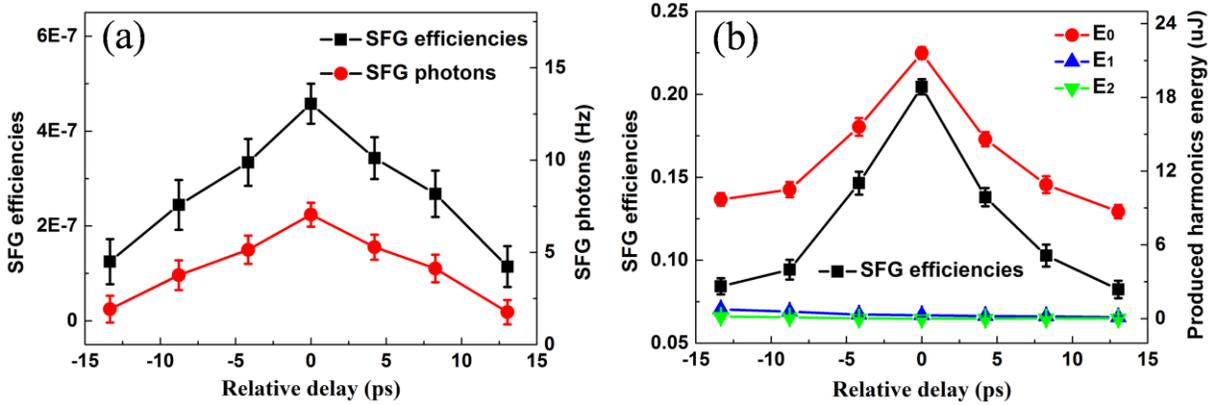

**Fig. 5** (color online). SFG photons and SFG efficiencies (a), SFG efficiencies and the energy of produced harmonics (b). The SFG efficiencies, SFG photons, the energy of produced harmonics, and error bars of them are accounted and the abscissa is a variable optical relative delay between the oppositely chirped laser pulse and the chirped laser pulse at the PPLN waveguide chip. The dark counts (3.5 Hz) are subtracted.